\documentclass[prl,aps,twocolumn,superscriptaddress,showpacs,amsmath,amssymb,floatfix]{revtex4}
                                                                                \usepackage{graphicx}

\newcommand{\be}{\begin{equation}}
\newcommand{\ee}{\end{equation}}
\newcommand{\wz}{\omega_{z}}
\newcommand{\rp}{r_{\perp}}

\begin{document}

\title{Sound propagation and oscillations of a superfluid  Fermi gas
in the presence of a 1D optical lattice}

\author{L.P. Pitaevskii}
\affiliation{Dipartimento di Fisica, Universita' di Trento and
BEC-INFM, 1-38050 Povo, Italy}
\affiliation{Kapitza Institute for Physical Problems, 117334 Moscow, Russia}

\author{S. Stringari}
\affiliation{Dipartimento di Fisica, Universita' di Trento and
BEC-INFM, 1-38050 Povo, Italy}
\date{\today}

\author{G. Orso}
\affiliation{Dipartimento di Fisica, Universita' di Trento and
BEC-INFM, 1-38050 Povo, Italy}

\begin{abstract}

We develop the hydrodynamic theory of Fermi superfluids in the presence of
a periodic potential. The relevant parameters governing the propagation of sound (compressibility and effective
mass) are calculated in the weakly interacting  BCS limit.
The conditions of stability of the superfluid motion with respect to creation
of elementary excitations are discussed. 
 We also evaluate the frequency of the center of mass oscillation 
when the superfluid  gas is additionally confined by a harmonic trap.

\end{abstract}
\maketitle

It is well known that  sound cannot propagate  in a 
degenerate noninteracting   Fermi gas. In fact
the sound velocity $c=v_F/\sqrt{3}$, calculated from the compressibility of 
the gas, is  smaller than the 
Fermi velocity $v_F$, so that  sound is unstable 
towards   decay into the  continuum of particle-hole excitations.
 In the presence of an attractive interaction, Cooper pairing opens
 a gap at the Fermi level. As a result, sound
waves with sufficiently low frequency can propagate 
 in a neutral  superfluid. An important question is to understand
how these modes behave when the system  is confined by a periodic potential.

The propagation of sound in Bose gases trapped by an optical lattice has 
been already the object of 
extensive theoretical work \cite{javanainen}. 
Experiments \cite{lens} on the collective oscillations  
have  confirmed that the low energy dynamic behaviour of a Bose-Einstein condensed  
gas is  well described by the hydrodynamic equations of superfluids 
\cite{meret},
where the coherent tunneling through the barriers generated by the periodic 
potential is taken into account
by a proper renormalization of the effective mass.

In this paper we generalize the hydrodynamic equations 
of Fermi superfluids in order to 
investigate the propagation of sound 
and the collective oscillations of a Fermi gas in the presence 
of a 1D periodic potential generated by a laser field. 

We consider an interacting two-component atomic   Fermi gas  
trapped by a potential $V_{ext}$ given by the sum 
of a 3D harmonic trap  and of a stationary 1D
 optical potential modulated along the $z$-axis: 
 \begin{equation}
 V_{ext} = V_{ho}  + s E_R\sin^2q_Bz, 
 \label{V}
 \end{equation}
 where $V_{ho}=m \left(\omega_{\perp}^2 \rp^2 + \omega_z^2z^2
 \right)/2$  and $E_R  = \hbar^2q_B^2/2m$ is the recoil energy. 
Here  $q_B$ is the 
Bragg momentum and $s$ is a dimensionless parameter providing  
the intensity of the laser beam.  The optical potential has periodicity 
 $d=\pi /q_B$  along the $z$-axis. 

The long wavelength dynamic behaviour of a neutral superfluid is properly
 described by  Landau's hydrodynamic theory. At $T=0$ the relevant 
variables are the density $n$ of particles and the superfluid velocity  
\begin{equation}\label{velo}
\mathbf{v}=\frac{\hbar }{2m}\nabla \phi, 
\end{equation}
where $m$ is the atomic mass and 
$\phi $ is the phase of the order parameter defined  
through the anomalous mean value $\langle
\Psi _{\uparrow }({\mathbf r})\Psi _{\downarrow }({\mathbf r})\rangle= 
|\langle \Psi _{\uparrow }({\mathbf r})\Psi _{\downarrow }({\mathbf r})\rangle|
e^{i\phi ({\bf r})}$. 

The hydrodynamic equations can be derived from a variational principle,
starting from an effective action describing the low energy
collective excitations over the ground state. In the weak coupling limit
these excitations are known as the Bogolubov-Anderson modes.
The Lagrangian of the system takes the general form 
\begin{equation}
L=\int d\mathbf r 
\Big[e(n,\mathbf v)+V_{ho}n+n\frac{\partial  }{\partial t}
\frac{\phi}{2}\Big]
\label{lag}  
\end{equation}
where $e(n,\mathbf v)$ is the energy per unit of volume.
By expanding $e(n,\mathbf v)$  up to the terms 
quadratic in $\mathbf v$ one finds
\be\label{giu}
e(n,\mathbf v)=e(n,0)+\frac{1}{2} n \mathbf v_{\perp}^2+
\frac{1}{2}\frac{m^2}{\tilde{m}}n v_z^2
\ee
where $\tilde{m
}$ ($\ge m$) is an effective mass  accounting for the 
increased inertia of the superfluid along the direction of the laser.
In general $\tilde{m}$ is density dependent. In Eq.(\ref{lag}) we have
assumed that the two spin species are equally populated $(n=2n_{\uparrow })$
. 
Inserting Eq.(\ref{giu}) in the Lagrangian (\ref{lag}) yields 
the  hydrodynamic  equations 
\begin{equation}
\frac{\partial n}{\partial t}+\partial _{x}(nv_{x})+\partial
_{y}(nv_{y})+\partial _{z}(\frac{m}{\tilde{m}}nv_{z})=0\,,  \label{mass}
\ee
\be
\frac{\partial }{\partial t}\mathbf{v} + \frac{1}{m}\nabla 
\Big(\mu (n)+V_{ho}+\frac{m}{2}v_x^2+ \frac{m}{2}v_y^2+
\frac{\partial}{\partial n}\big(\frac{m}{\tilde m}n\big)  \frac{m}{2} v_z^2\Big) =0,
\label{vel}
\end{equation}
where $\mu (n)=\partial e/\partial n$ is the chemical potential.
Generalization to include 2D or 3D optical lattices is trivial.

Let us first consider the propagation of sound in the absence of the 
harmonic potential ($V_{ho}=0$). We will include later the 
harmonic trap  to investigate the 
oscillations of the cloud. 
If $V_{ho}=0$ Eqs. (\ref{mass}) and (\ref{vel}) admit linearized  
solutions around the equilibrium density $n_0$ of the form
  $n=n_0+\delta n $, with
$\delta n \sim e^{iq z-i\omega t}$ and analogously for $v_z$. 
This permits to derive  the phonon dispersion $\omega=c_z q$ along the 
direction of the optical lattice, with the sound velocity  given by 
\be
\label{sound}
c_z=\sqrt{\frac{n}{\tilde m} \frac{\partial \mu}{\partial n}}.
\ee
The presence of the periodic lattice enters Eq.(\ref{sound})
through both the renormalization of $\tilde m$ and 
the change of the inverse compressibility $n\partial \mu /\partial n$.
Analogously, the sound velocity along the transverse 
directions, where the gas is free, is given by 
$c_{\perp}^2=(n/m)\partial \mu /\partial n$.   

Equations  (\ref{mass}-\ref{sound}) are very general and apply also 
to strongly interacting superfluids. They permit to 
calculate the low energy dynamics of the system once the equation of state 
and the effective mass are known. In the following 
we will apply these equations to  a dilute Fermi gas interacting 
with negative scattering length (BCS limit). In this case
the internal energy is fixed by the quantum pressure and therefore 
the chemical potential is equal to the Fermi energy $(\mu=\epsilon_F)$.
For simplicity we consider values of the Fermi energy
such that only the lowest Bloch band is populated at zero temperature. 
The single particle energy spectrum can then be written as 
$\epsilon(\mathbf k)=\mathbf k_\perp^2/2m+\epsilon(k_z)$ where
$\epsilon(k_z)$ is the dispersion relation of the lowest Bloch band.
The energy of the system at rest is given by $E=V\int \epsilon \left( 
\mathbf{k}{}\right) n_{\mathbf{k}}d\mathbf{k} /(2\pi\hbar)^3,$ where 
$n_\mathbf{k}=2\Theta(\epsilon_F-\epsilon(\mathbf k))$ is the quasi-momentum 
distribution of the non-interacting gas \cite{nk}, $\Theta(x)$ 
being the usual step function, $V$ is the volume of the system and 
the integration over $k_{z}$ is restricted to the first Brillouin 
zone $-\pi \hbar /d\leq k_{z}<\pi \hbar /d$. 
 The Fermi energy $\epsilon_F$  is related to the atomic density by
the normalization condition $n=\int n_{\mathbf k}d\mathbf k/(2\pi\hbar)^3$.  
 
Let us first assume that the gas moves with constant superfluid 
velocity $\mathbf{v}$. The order parameter  acquires the phase
$\phi =2m\mathbf{v\cdot r} /\hbar $ (see Eq.(\ref{velo})) and,
correspondingly, the field  operator $\hat \Psi_\sigma $ transforms according 
to  
\begin{equation}
\hat \Psi_{\sigma}(\mathbf r) \rightarrow \hat \Psi_{\sigma}(\mathbf r) e^{im\mathbf{v\cdot r}/\hbar }.  \label{gauge}
\end{equation}
Equation (\ref{gauge}) is a gauge transformation which 
does not change the energy levels, but changes their classification. 
In particular the Bloch state of quasi-momentum $\mathbf{k}$ 
is mapped into the state $\mathbf{k}+m\mathbf{v}$ (and 
hence $\epsilon(\mathbf{k})$ into
$\epsilon(\mathbf{k}+m\mathbf{v})$). The energy density of the moving system 
can therefore be written as:
\begin{equation}
e(n,\mathbf v)=\int \epsilon \left( \mathbf{k}+m\mathbf{v} \right) n_{\mathbf{k}}\frac{d
\mathbf{k}}{\left( 2\pi \hbar \right) ^{3}}  \label{delta}
\end{equation}
where the density dependence comes from the quasi-momentum 
distribution $n_\mathbf{k}$.
We can now expand Eq.(\ref{delta}) in powers of $\mathbf{v}$. The linear term
vanishes by symmetry as required by the stability of the ground state. The
first non vanishing contribution is the quadratic, diamagnetic term.
Comparison  with Eq.({\ref{giu}) then permits to find, after an 
integration by parts, the useful expression  
\begin{equation}
\frac{1}{\tilde{m}}=\frac{\int (\partial \epsilon /\partial k_{z})^{2}\delta
(\epsilon _{F}-\epsilon(\mathbf k))d\mathbf k /(2\pi\hbar)^3}{n/2}  
\label{mm}
\end{equation}
for the effective mass, holding for a dilute superfluid Fermi gas 
at zero temperature.
 From Eqs (\ref{delta}) and (\ref{mm}), we also see that 
the current density along 
}$z$ in the equation of continuity (\ref{mass}) corresponds to $
j_{z}=(1/mV)\delta E/\delta v_{z}$.

From Eq.(\ref{mm})  we see that only the states near the Fermi surface 
contribute to the effective mass. In general $\tilde m$ 
depends on the Fermi energy or, equivalently, on the density.
An important exception occurs when the Bloch band is almost empty.  
In this case the occupied states obey an approximate quadratic
dispersion $\epsilon(k_z)=k_z^2/2m^*$, where $m^*$ depends
only on the laser intensity. From Eq.(\ref{mm}) we then find
$\tilde m=m^*$. This coincides with the result for the effective mass 
holding for a 
dilute Bose-Einstein condensed gas in the same optical lattice \cite{book}. 
Equation (\ref{mm}) also shows that the integration along the transverse
directions plays a crucial role in keeping $\tilde m$ finite as 
the Fermi energy crosses the Bloch bandwith. In fact if we neglect the 
radial dispersion and consider a pure 1D system, the effective mass
(\ref{mm}) would actually diverge as the band is completely filled. 

A second crucial ingredient needed to calculate the velocity of sound 
(see Eq.(\ref{sound})) is the compressibility 
$(n\partial \mu /\partial n)^{-1}$. 
In a dilute Fermi gas at zero temperature this is simply calculated  
in terms of the density of states as
\begin{equation}
\Big (n\frac{\partial \mu}{\partial n} \Big)^{-1}=
\frac{\int \delta
(\epsilon _{F}-\epsilon(\mathbf k))d\mathbf k /(2\pi\hbar)^3}{n/2}.
\label{compressibility}
\end{equation}
The optical lattice modifies the compressibility with respect to
the value $ (3\pi^2)^{2/3}\hbar^2 n^{2/3}/3m$ holding for a uniform gas.
By inserting Eqs (\ref{mm}) and (\ref{compressibility}) into 
Eq.(\ref{sound}),  we  find the result
\be\label{speed}
c_z^2=\frac{\int (\partial \epsilon/\partial k_z)^2  
\delta(\epsilon_F-\epsilon(\mathbf k))d\mathbf k }
{\int \delta(\epsilon_F-\epsilon(\mathbf k))d\mathbf k}
\ee
for $c_z^2$, which can be regarded as the average of 
the square of the group velocity
$\partial \epsilon/\partial k_z$ over the Fermi surface. 
Again it should be noted that, due to the free dispersion in the 
transverse direction, the sound velocity
(\ref{speed}) remains finite as $\epsilon_F$ crosses the bandwidth. 
In a pure 1D system, Eq.(\ref{speed}) would predict $c_z=0$ 
when the band is completely filled,
 since $\partial \epsilon/\partial k_z$ vanishes at the edge of 
the Brillouin zone.

Let us point out that Eq.(\ref{speed}) has been derived starting
from the hydrodynamic equations but the same result can be obtained
 by generalizing BCS theory through the inclusion of  the interaction between Bogolubov quasi-particles.
This can be achieved, for example, using
time dependent Hartree-Fock theory \cite{schrieffer} or the Random Phase 
Approximation (RPA) \cite{anderson}.

If the  laser intensity is sufficiently large one can  work
in the tight-binding approximation where the dispersion
of the lowest Bloch band takes the simple form  
$\epsilon(k_z)=\delta\big(1-\cos(k_z d/\hbar)\big)$, with  
$\delta$ proportional to the tunneling rate between two consecutive wells.
In this case the equation of state can be calculated analytically. 
For values of the Fermi energy above the bandwidth $(\epsilon_F >2\delta)$ 
one finds $n=m(\epsilon_F -\delta)/\pi\hbar^2 d$ and hence
$dn/d\mu=m/\pi\hbar^2d $.
For smaller values  ($\epsilon_F \le 2\delta$) one instead finds    
$n= \big(2 \sqrt{y(1-y)}+(2y-1)\arccos(1-2y)\big )m\delta /\pi^2\hbar^2 d$,
yielding $dn/d\mu=(m/\pi^2 \hbar^2 d)\arccos(1-2y)$,
 where $y=\epsilon_F/2\delta$. 

The effective mass (\ref{mm}) increases as the density of particles
increases. For values of $\epsilon_F < 2\delta$ one finds
$\tilde m=(2 \pi^2 \hbar^4 n/\delta^2 d m) [2(2y-1)\sqrt{y(1-y)}+\arccos(1-2y)]^{-1}$ and in the limit $y\ll 1$, corresponding to 
$\epsilon_F \ll 2\delta$, $\tilde m$ approaches
the density independent value $m^*=\hbar^2/\delta d^2$. Conversely, 
for $\epsilon_F >2\delta$ the effective mass scales linearly with the
density: $\tilde m=2\hbar^4 \pi n/m\delta^2 d$.  This result can be 
obtained in a BCS approach, following Ref.\cite{tanaka}. 

Let us now use Eq.(\ref{sound}) to evaluate the behaviour of  
the sound velocity along the direction of the optical confinement.  
In the limit $\epsilon_F >2\delta$
one finds the density independent value $c_z^2=\delta^2 d^2/2\hbar^2$.
This is simply understood by noticing that 
$\partial \epsilon/\partial k_z=(\delta d/\hbar)\sin(k_z d/\hbar)$
and that the average of $\sin^2(k_z d/\hbar)$ over the 
first Brillouin zone gives 
the factor $1/2$. 
For $\epsilon_F<2\delta$, 
the longitudinal sound velocity is related to the Fermi energy by 
the expression
\be
\label{cz}
c_z^2=\big (\frac{\delta d}{\hbar}\big )^2 \Big [\frac{(2y-1)\sqrt{y(1-y)}}
{\arccos (1-2y)}+\frac{1}{2}\Big].
\ee
In particular, for $y \ll 1$, corresponding to 
$\epsilon_F \ll 2\delta$, Eq.(\ref{cz}) reduces to 
$c_z^2=2\epsilon_F/3m^*$. It is important to notice that the continuous 
hydrodynamic approach along the direction of the laser field 
requires that the phase $\phi$ of the order parameter varies slowly
between neighboring wells. Taking into account that the energy of
the phonons must be small compared to the superfluid gap $\Delta$, we see
that our theory is valid provided 
$q_z \ll \textrm{min} (\Delta/\hbar c_z,d^{-1}) $ where 
$q_z$ is the wave vector of the sound excitation.  

The propagation of sound in the transverse direction is affected by the 
presence of the optical lattice only through the change of the compressibility.
For values of the density corresponding to $\epsilon_F >2\delta$, one
finds $c_\perp^2=(\epsilon_F-\delta)/m$. For smaller values of 
$\epsilon_F$ one finds
$c_\perp^2=\pi^2 \hbar^2 d n/m^2\arccos(1-2y)$, yielding 
$c_\perp^2=2\epsilon_F/3 m$ in the limit $\epsilon_F \ll 2\delta$.
In the case of transverse sound propagation, the hydrodynamic theory
is valid under the usual assumptions $q_\perp \ll \Delta/\hbar c_\perp$.  

The hydrodynamic equations (\ref{mass},\ref{vel}) can be generalized 
to include finite velocity of the gas with respect to the optical lattice.  
This is important in order to study the conditions of stability of the 
moving fluid with respect to the creation of elementary excitations.
This problem has been recently addressed in the case of 
Bose-Einstein condensates, both from the theoretical \cite{chiara} 
and experimental \cite{inguscio} point of view.

Starting from the general form (\ref{lag}) for the Lagrangian 
and setting $V_{ho}=0$, one finds
\begin{eqnarray}\label{eq1}
\frac{\partial n}{\partial t}+\nabla \mathbf j(n,\mathbf v)=0,\\
m \frac{\partial \mathbf v}{\partial t}+\nabla \mu(n,\mathbf v)=0,\label{eq2}
\end{eqnarray}
where $\mathbf j=(1/m) \partial e /\partial \mathbf v$, $\mu=\partial e /\partial n$ are, respectively, the current and the chemical potential of the gas
while $e(n,\mathbf v)$ is defined in Eq.(\ref{delta}).

Linearizing Eq.s (\ref{eq1},\ref{eq2}) with 
$n(\mathbf r)=n_0+\delta n(z)$ and 
$\mathbf v(\mathbf r)=(k/m+\delta v(z))\bf \hat{z}$ and
looking for solutions of the form $\delta n, \delta 
v\sim e^{i (qz-\omega t)}$, we find the dispersion relation
\be\label{stab}
\omega=\frac{\partial^2 \epsilon}{\partial n \partial k}q+
\sqrt{\frac{\partial^2 \epsilon}{\partial n^2}
\frac{\partial^2 \epsilon}{\partial k^2}}|q|.
\ee
Dynamical instability occurs when the argument in the square root 
of Eq.(\ref{stab}) becomes negative. This leads to complex frequency
and corresponds to an exponential growth of excitations wich destabilize
the superfluid.
Energetic instability instead takes place when $\omega$ becomes negative,
 i.e. when the first term in Eq.(\ref{stab}) (accounting for the Doppler shift) is in modulus larger than the second one.

In tight-binding limit the energy density (\ref{delta}) takes the 
simple form
\be\label{first}
\epsilon(n,k)=\epsilon(n,0)+\frac{n}{\tilde m}\big(\frac{\hbar}{d}\big)^2 \big(1-\cos[\frac{kd}{\hbar}]\big) 
\ee
where $\tilde m$ is defined in Eq.(\ref{mm}). Note that in the limit 
of small $k$ Eq.(\ref{first}) reduces to Eq.(\ref{giu}) with 
$\mathbf v_\perp=0$.
By inserting Eq.(\ref{first}) into Eq.(\ref{stab}) we find
\be\label{re1}
\omega=\frac{\partial}{\partial n}\Big(\frac{n}{\tilde m}\Big)\frac{\hbar}{d}
\sin[\frac{kd}{\hbar}]q+\sqrt{\frac{\partial^2 \epsilon}{\partial n^2} \frac{n}{\tilde m}\cos[\frac{kd}{\hbar}]}|q|. 
\ee
 We see from  Eq.(\ref{re1}) that in tight-binding 
limit the instability opens at $k=\pi \hbar/2 d$ 
similar to what happens in Bose-Einstein condensates \cite{chiara}.
One should notice that results (\ref{first},\ref{re1}),
being derived from hydrodynamic theory, holds only for small 
values of $q$ \cite{not}.
There are however no restrictions on the value of the stationary
velocity $k/m$ of the fluid with respect to the lattice. 

For $\epsilon_F \ll 2\delta$, where $\tilde m= m^*$, the Doppler term is 
exactly the same as for Bose condensates. For higher values of the 
density, however, this term becomes smaller and vanishes identically 
for $\epsilon_F>2\delta$, where $\tilde m$ is linear 
in the density $n$. In this limit  the current becomes density independent
while the chemical potential no longer depends on the velocity of the fluid. 
   
So far we have discussed the  behaviour of a superfluid gas
confined by  a  periodic potential.  In the presence of the additional 
harmonic potential $V_{ho}$ the low energy oscillations
exhibit new features.
Of particular importance   is  the center of mass
oscillation along the direction of the optical lattice. 
In the absence of the periodic potential, this mode would oscillate
 exactly at the frequency of the harmonic trap. The presence of the periodic 
 potential gives rise to new interesting features. 
 In a recent work \cite{luca} it has been shown  that in 
a non-interacting Fermi gas with Fermi energy $\epsilon_F$ 
larger than the Bloch bandwith $2 \delta$,  the center of mass 
cannot oscillate around the bottom of 
the trap, but remains localized at one side of the harmonic field.  
Inclusion of collisions  in a two-component gas favours the relaxation 
towards the equilibrium configuration. However,
collisions cannot restore the propagation of sound in the hydrodynamic 
regime as happens in the absence of the periodic potential. 
In fact, it was  concluded in Ref.\cite{umklapp} that, under the condition 
$\epsilon_F >2 \delta$, the center of mass oscillations of
a non superfluid gas are overdamped in the collisional regime, 
due to umklapp processes.

As recently suggested by Wouters et al. \cite{wouters}, 
in the superfluid   regime
the center of mass can instead oscillate and 
the corresponding frequency can be calculated starting from the 
hydrodynamic equations (\ref{mass}) and (\ref{vel}).
We will  make the natural ansatz 
\be\label{ansatz}
\phi=\alpha(t)z,\;\; n=n_0(z-a(t),\mathbf r_\perp),
\ee
for the phase and for the density distribution 
where $\alpha$ and $a$ are functions of time 
and $n_0(\mathbf r)=2\int \Theta (\epsilon_F(\mathbf r)-\epsilon(\mathbf k)) d\mathbf k/(2\pi\hbar)^3$ is the equilibrium density evaluated in the local density approximation with 
$\epsilon_F(\mathbf r)=\epsilon_F-V_{ho}(\mathbf r)$.
This ansatz  corresponds to a rigid shift of the density 
in coordinate space accompained by a uniform 
velocity field $v_z=\hbar \alpha/2 m$.
By  inserting the ansatz (\ref{ansatz}) into the Lagrangian (\ref{lag}) and
retaining only the lowest order (quadratic) terms in the 
functions $a$ and $\alpha$,  one can recast (\ref{lag}) in the form  
$L(\alpha,\dot{\alpha},a) \propto m\omega_z^2 a^2/2+\alpha^2/8 m_{CM}+
a\dot{\alpha}/2$, where we have introduced the averaged effective 
mass \cite{notecollisionalHD}
\begin{eqnarray}
\frac{1}{m_{CM}}=
\frac{1}{N}\int  \frac{1}{ \tilde{m}}n_0(\mathbf r)d \mathbf r \nonumber\\
= \frac{2}{N} \int (\partial \epsilon /\partial k_{z})^{2}\delta
(\epsilon _{F}(\mathbf r)-\epsilon(\mathbf k))d\mathbf k d\mathbf r/(2\pi\hbar)^3 \label{super}
\end{eqnarray}  
and $N=\int n_0(\mathbf r)d \mathbf r$ is the total number of particles.
The above Lagrangian describes a simple 
harmonic oscillator with frequency 
\be\label{freq}
\omega_{CM}=\omega_z \sqrt{\frac{m}{ m_{CM}}}.
\ee
Result (\ref{freq}) for the frequency of the center of mass oscillation can be also derived microscopically
using a sum rule approach,  based on the ratio 
\begin{equation}
(\hbar\omega)^2 = \frac{m_1}{ m_{-1}}
\label{sumrules}
\end{equation}
between the energy weighted  and inverse energy weighted  moments 
\cite{book}  relative to the dipole operator $D= \sum_{j=1}^N z_j$.
The energy weighted moment can be calculated using the general expression
$m_1= (1/2)\langle[D,[H,D]]\rangle$ where 
$\langle\rangle $ is the average over the ground state
 and  the effective Hamiltonian $H=\sum_j\epsilon({\bf k}_j) + \sum_jV_{ho}({\bf r}_j)
+ \sum_{j<\ell}V_{2-body}(\mid {\bf r}_k-{\bf r}_{\ell}\mid)$ 
is the projection of the full Hamiltonian into the lowest Bloch band. One  finds 
$m_1= (1/2)\langle \sum_{j=1}^N\partial^2 \epsilon/\partial k_{z_j}^2\rangle $, which is the analog of the f-sum rule 
\cite{fsumrule}. On the other hand the inverse energy weighted moment 
is easily related  to
the  dipole static polarizability $\alpha_D$, fixed by the harmonic 
potential: $m_{-1}=(1/2)\alpha_D =
N/(2m\omega_z^2)$. By evaluating the average  $\langle \sum_{j=1}^N\partial^2 \epsilon/\partial k_{z_j}^2\rangle$ in the limit of a weakly interacting gas  and inserting the results for $m_1$ and $m_{-1}$ into Eq.(\ref{sumrules}), one  recovers result (\ref{freq}) 
for the  frequency of the center of mass oscillation. 

\begin{figure}
\begin{center}
\includegraphics[height=8cm,angle=270]{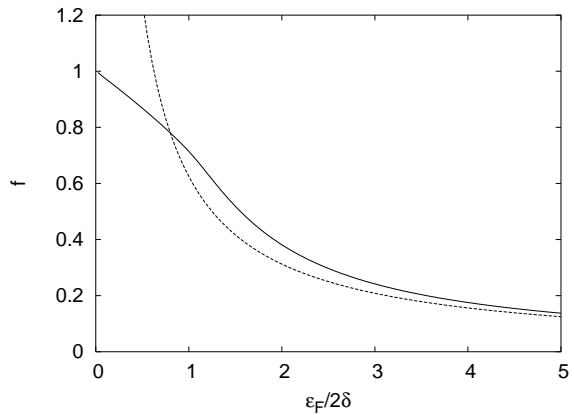}
\caption{Function $f$ [See Eq.(\ref{final})] versus 
the parameter $y=\epsilon_F/2\delta$ (solid line). The asymptotic
limit $5/8y$ is also shown (dashed line).}
\label{frequ}
\end{center}
\end{figure}

Equations (\ref{super}), (\ref{freq}) hold for any value of the laser 
intensity provided the system remains superfluid.  
In the tight binding limit the formalism becomes particularly 
simple. In fact in this case the effective mass (\ref{super}), 
can be conveniently written as $1/m_{CM}=f/m^*$, where 
\be\label{final}
f=\frac{5}{2}\frac{\int 
h(x)^{3/2}  \sin^2 x \Theta[h(x)]dx} 
{\int h(x)^{5/2} \Theta[h(x)]dx}
\ee 
is a dimensionless function of the ratio $\epsilon_F/2\delta$ and
the dependence of $m^*=\hbar^2/\delta d^2$ on the laser intensity $s$
 is given, for example, in Ref.\cite{book}.
The function $h(x)$ is defined by
$h(x)=\epsilon_F/\delta-1+\cos x$. Here $\epsilon_F$ is the Fermi energy 
which is related to the free value 
$\epsilon_F^0={(3N)}^{1/3}\hbar ({\omega_\perp}^2\wz)^{1/3}$
evaluated in the absence of the periodic potential, by the equation 
\be\label{number}
(\epsilon_F^0)^3=\frac{16}{5 \pi^2}\Big(\frac{E_R}{\delta}\Big)^{1/2}
\delta^{3}
 \int_{-\pi}^{\pi}
h(x)^{5/2} \Theta[h(x)]dx,
\ee
which follows from the normalization condition 
$N=\int n_0(\mathbf r)d \mathbf r$.
 
In Fig.\ref{frequ} we plot $f$  as a function of the  parameter
$y=\epsilon_F/2\delta$.
In the limit $\epsilon_F \ll 2\delta$, 
$\tilde m$ does not depend on the density and one finds
$f=1$ and  $m_{CM}=m^*=\hbar^2 /\delta d^2$. 
For higher values of the density, the frequency of the oscillation 
becomes smaller because $ m_{CM}$ increases with the density.  
In the limit $\epsilon_F \gg 2\delta$, we find the asymptotic
behavior $f=5\delta/4\epsilon_F$ (see dashed line in Fig.\ref{frequ}) and  
$m_{CM}= 4\hbar^2 \epsilon_F/5 d^2 \delta^2$.

As an example, we consider a two-component gas of $N=10^5$ 
potassium $(^{40}K)$ 
atoms with trap frequencies $\omega_\perp=2 \pi \cdot 275$Hz and
$\wz=2 \pi \cdot 24 $Hz, corresponding to $\epsilon^0_F/k_B= 390nK$.
For the optical lattice we assume 
$s=5$ and periodicity $d=400$nm, corresponding to  
$E_R=9.2 \delta=374 nK$.
From Eq.(\ref{number}) one finds 
$\epsilon_F = 0.85 \epsilon_F^0$
and from the value $\epsilon_F/2\delta=4.1$, one finds $f(4.1)=0.17$,
yielding $m_{CM}/m=10.94$. Eq.(\ref{freq}) predicts a significant reduction 
($\omega_{CM}=0.30 \omega_z$) 
of the frequency of the center of mass oscillation. Notice indeed that 
this reduction 
is larger than in the corresponding case  of an oscillating  
Bose-Einstein condensate,  where 
$\omega_{CM}=\sqrt{m/m^*}\omega_z = 0.73\omega_z$.

In conclusion we have investigated the dynamic behaviour of a superfluid 
Fermi gas in the presence of a 1D optical lattice. We have calculated the 
velocity of sound 
and studied the conditions of stability (both energetical and dynamical)
against the creation of elementary excitations when the fluid moves
with respect to the lattice.  
Moreover, we have calculated the frequency of the center 
of mass oscillation (in the BCS limit) when the gas is further confined 
by a harmonic trap. 
 Our results might be useful for the identification of the superfluid phase of 
interacting Fermi gases. Near a Feshbach resonance, where the scattering length becomes very large, one still expects
the sytem to remain superfluid for moderate intensities of the laser field and, consequently, to exhibit undamped center of mass oscillations in the presence of harmonic trapping.
In this case, however, the evaluation of the effective mass cannot be based on Eqs.(\ref{mm},\ref{super}) and requires a more complete
many-body calculation.

This work was supported by the Ministero dell'Istruzione, 
dell'Universita' e della Ricerca (M.I.U.R.).


\begin{thebibliography}{15}
\bibitem{nk} Since in a weakly interacting gas the  Fermi energy
is much larger than the superfluid gap, 
the momentum distribution in the superfluid differs very little from the noninteracting  value $n_k$.
\bibitem{not}In the case of Bose condensates a generalization of the 
above results to arbitrary values of $q$ can be obtained in the framework 
of the Gross-Pitaevskii theory, see Ref. \cite{chiara}.
\bibitem{notecollisionalHD} Notice that result (\ref{super}) for the effective mass  of the superfluid trapped gas 
differs from the result $m_{CM}= (2/N) \int k^2_z\delta
(\epsilon _{F}(\mathbf r)-\epsilon(\mathbf k))d\mathbf k d\mathbf r/(2\pi\hbar)^3$  holding in the collisional regime  of a degenerate Fermi 
gas \cite{umklapp}. The two results coincide only if the dispersion
$\epsilon(k_z)$ is quadratic.
\end{thebibliography}
\end{document}